\documentclass[10pt,english,journal]{IEEEtran}
\usepackage[T1]{fontenc}
\usepackage[latin9]{inputenc}
\usepackage{geometry}
\usepackage{amsmath}
\usepackage{xpatch}
\usepackage{amssymb}
\usepackage{esint}
\usepackage{mathtools}
\usepackage{amsthm}
\usepackage{nicefrac}
\usepackage{bigints}
\usepackage{mathtools}
\usepackage{blkarray, bigstrut}
\usepackage{physics}
\usepackage{calligra}
\usepackage{graphicx}
\usepackage{epstopdf}
\usepackage{dsfont}
\usepackage{array,ragged2e}
\usepackage{enumitem}
\usepackage{etoolbox}
\usepackage{babel}
\usepackage[nopar]{lipsum}
\usepackage{psfrag}
\usepackage[ruled,lined,linesnumbered]{algorithm2e}
\usepackage{algorithmic}
\usepackage{algorithm2e}
\usepackage{balance}
\usepackage{float}
\usepackage{hyperref}

\usepackage{subcaption}

\SetKw{KwBy}{by}

\makeatletter
\newcommand{\removelatexerror}{\let\@latex@error\@gobble}
\makeatother

\graphicspath{ {Figures/} }

\xpatchcmd{\proof}{\hskip\labelsep}{\hskip5\labelsep}{}{}  
\makeatletter
\xpatchcmd{\proof}{\@addpunct{.}}{\@addpunct{:}}{}{}
\makeatother

\renewcommand\[{\begin{equation}}
\renewcommand\]{\end{equation}} 
\usepackage{listings}
\usepackage{fancyvrb}
\usepackage{framed}

\usepackage{courier}
\usepackage[usenames,dvipsnames,table]{xcolor}

\definecolor{dkgreen}{rgb}{0,0.3,0}
\definecolor{gray}{rgb}{0.5,0.5,0.5}

\makeatletter
\newcommand*{\rom}[1]{\expandafter\@slowromancap\romannumeral #1@}
\makeatother

\usepackage{siunitx}
\usepackage{tabu}
\usepackage{booktabs}
\usepackage{multirow}
\usepackage{capt-of}
\usepackage{array}
\usepackage{arydshln}
\newcommand{\comment}[1]{}

\begin{document}

\title{



SliceOps: Explainable MLOps for Streamlined Automation-Native 6G Networks

}

\author{
Farhad~Rezazadeh,~\IEEEmembership{Student~Member,~IEEE}, 
Hatim~Chergui,~\IEEEmembership{Senior~Member,~IEEE}, 
Luis~Alonso,~\IEEEmembership{Senior~Member,~IEEE}, and~Christos~Verikoukis,~\IEEEmembership{Senior~Member,~IEEE}


\IEEEcompsocitemizethanks{\IEEEcompsocthanksitem F. Rezazadeh is with the Telecommunications Technological Center of Catalonia (CTTC) and  Technical University of Catalonia (UPC), 08860 Castelldefels, Spain (e-mail: frezazadeh@cttc.es).}
\IEEEcompsocitemizethanks{\IEEEcompsocthanksitem H. Chergui is with the i2CAT Foundation, 08034 Barcelona, Spain (e-mail:  chergui@ieee.org).}
\IEEEcompsocitemizethanks{\IEEEcompsocthanksitem Luis Alonso is with the Technical University of Catalonia (UPC), 08860 Castelldefels, Spain (e-mail: luisg@tsc.upc.edu).}
\IEEEcompsocitemizethanks{\IEEEcompsocthanksitem C. Verikoukis is with the University of Patras, ATHENA/ISI, Greece, and IQUADRAT Informatica, Barcelona 08006, Spain (e-mail: chverik@gmail.com).}
}

\maketitle

\begin{abstract}
Sixth-generation (6G) network slicing is the backbone of future communications systems. It inaugurates the era of extreme ultra-reliable and low-latency communication (xURLLC) and pervades the digitalization of the various vertical immersive use cases. Since 6G inherently underpins artificial intelligence (AI), we propose a systematic and standalone slice termed SliceOps that is natively embedded in the 6G architecture, which gathers and manages the whole AI lifecycle through monitoring, re-training, and deploying the machine learning (ML) models as a service for the 6G slices. By leveraging machine learning operations (MLOps) in conjunction with eXplainable AI (XAI), SliceOps strives to cope with the opaqueness of black-box AI using explanation-guided reinforcement learning (XRL) to fulfill transparency, trustworthiness, and interpretability in the network slicing ecosystem. This article starts by elaborating on the architectural and algorithmic aspects of SliceOps. Then, the deployed cloud-native SliceOps working is exemplified via a latency-aware resource allocation problem. The deep RL (DRL)-based SliceOps agents within slices provide AI services aiming to allocate optimal radio resources and impede service quality degradation. Simulation results demonstrate the effectiveness of SliceOps-driven slicing. The article discusses afterward the SliceOps challenges and limitations. Finally, the key open research directions corresponding to the proposed approach are identified.

\end{abstract}


\section{Introduction}
\IEEEPARstart{6}{G} slicing is envisioned as a disruptive technology that intelligently supports various verticals with different quality of service (QoS) requirements. Such massive slicing is viewed as a new paradigm in 6G network that supports numerous slices with micro or macro services. Consequently, the complexity of automated management and orchestration (MANO) operations would arise dramatically. The tendency towards fully automated MANO in beyond 5G/6G has spurred intensive research interest in applying AI and ML as an ideal solution for various nonlinear problems. Notably, novel practices are required to deploy ML solutions into production and providing a trustworthy and actionable AI-driven slicing environment. In this intent, the ML model deployed in production should inescapably undergo the combination of development and operations (DevOps) and ML Operations (MLOps). On the other hand, following the technical report of the European Commission on "Ethics guidelines for trustworthy AI" \cite{XAI-Eth}, the AI solutions should pursue trustworthiness. Indeed, the XAI approach scrutinizes each feature and its impact on the output of the AI model, enabling to observe the factors that either positively or negatively impact the AI model prediction. 

In this regard, XRL is viewed as a responsible and trustful ML approach that can be combined with MLOps lifecycle. Indeed, in the reinforcement learning (RL) method, the agent generates the corresponding dataset on the fly by interacting with the network. This method is an evaluative and feedback-based learning to optimize the accumulated long-term reward. Despite the promising results and performance, there is a concern about the essence of deep neural network (DNN) that are deemed as opaque models. This issue is exacerbated when high reliability and security play a vital role in realistic network scenario and impedes users from trusting the trained agents and predicted results in the network slicing ecosystem. Moreover, the conflict-prone nature of state-action can be highly crippling the promising RL solutions in automated network slicing. Motivated by explanation-guided learning (EGL), we consider an intrinsic interpretability approach in the training phase of the proposed SliceOps agent. In this intent, XAI can assist DRL to extract more relative state-action pairs where it explains which state or input of agent has the most positive impact on action or decision.

To incorporate these principles into a single design while ensuring a separation of concerns, this paper introduces \emph{SliceOps}, an XAI-empowered MLOps framework that is natively embedded in the 6G network architecture as a standalone slice. To illustrate its operation, we consider a latency-aware resource allocation problem where each slice registers to the corresponding SliceOps instance which provides AI services via a SliceOps agent. The main goal is to allocate optimal radio resources to the slices while minimizing the latency to meet the service level agreements (SLAs). The following contributions are presented in this paper:
\begin{itemize}

\item We introduce the architecture of SliceOps, where the explainable ML operations are gathered in a standalone slice providing AI services to the rest of the slices. This continuous delivery (CD) and continuous integration (CI) of ML models enhances reliability and interpretability while quickly deploying AI models in the network with higher consistency.
\item As a use case of SliceOps, a RAN resource allocation problem is defined, aiming at reducing SLA violations.
\item To solve this problem, SliceOps agents are proposed, which are based on a novel explanation-guided DRL (XRL) scheme, that is assisted with shapley additive explanations (SHAP) importance values and an entropy mapper to guide the agent in reducing uncertainty in its actions across various network states.
\item The AI and network analysis demonstrate the superiority and faithfulness of the proposed explanation-guided DRL approach compared to the RL baseline.
\end{itemize}

For the sake of exploring the aspects mentioned above, the article starts by highlighting the background of this work. 
Following that, we discuss the explainability in RL and the interactions between the different components of the SliceOps workflow. We illustrate the performance analysis and effectiveness of the proposed SliceOps-driven resource allocation. Also, we reveal the challenges and limitations of the proposed AI-native slice. Finally, the open research directions for implementation of 6G SliceOps are highlighted and concludes the article.

\section{Research and standardization work relevant to MLOps}

The authors in \cite{Raj-mlops} developed an edge MLOps framework for automating and more efficient artificial intelligence of things (AIoT) operations and decision-making. The proposed framework aims to operationalize the CD and CI of ML models to the nodes as an essential part of DevOps. 
Samaras \emph{et al.} \cite{Samaras-mlops} proposed a cloud-native MLOps automation platform for automated and optimized network slice lifecycle management (LCM). 
This event-based framework is a zero-touch MLOps platform that automatically analyzes its accuracy performance and adapts itself automatically through the training job. In \cite{Zaidi_mlops}, the authors presented an ML-as-a-Service (MLaaS) approach for 5G IoT based on a TinyMLaaS (TMLaaS) architecture. 
They leveraged an MLOps framework for unifying ML systems to implement, deploy, and maintain operations. The authors in \cite{Brik_mlops} described two levels of MLOps in open radio access network (O-RAN). They considered a deep learning (DL)-based MLOps and leveraged the DevOps principles to ease ML system development (Dev) and ML system operations (Ops) for O-RAN automation. Correspondingly, \cite{Li-mlops} proposed an MLOps lifecycle based on RL aiming to automate and reproducible model development process in the O-RAN deployment. The proposed scheme introduces principles and practices for developing data-derived optimal decision-making strategies 
Tsourdinis \emph{et al.} \cite{Tsourdinis-mlops} developed a service-aware dynamic slice allocation scheme. The proposed AI/ML unit in the pipeline follows an MLOps-based distributed ML architecture. Recently, reward shaping~\cite{r-shaping} and attention mechanisms~\cite{att-mech} have emerged as prominent techniques in enhancing the explainability of DRL methods. Reward shaping entails altering the rewards that an agent receives to offer supplementary guidance to the agent. Moreover, attention mechanisms empower the agent to concentrate on the pertinent network states and inputs. It presents the importance of each feature or state in state-action pairs for effective decision-making. 
\begin{figure*}[h]
\centering
\includegraphics[scale=0.73]{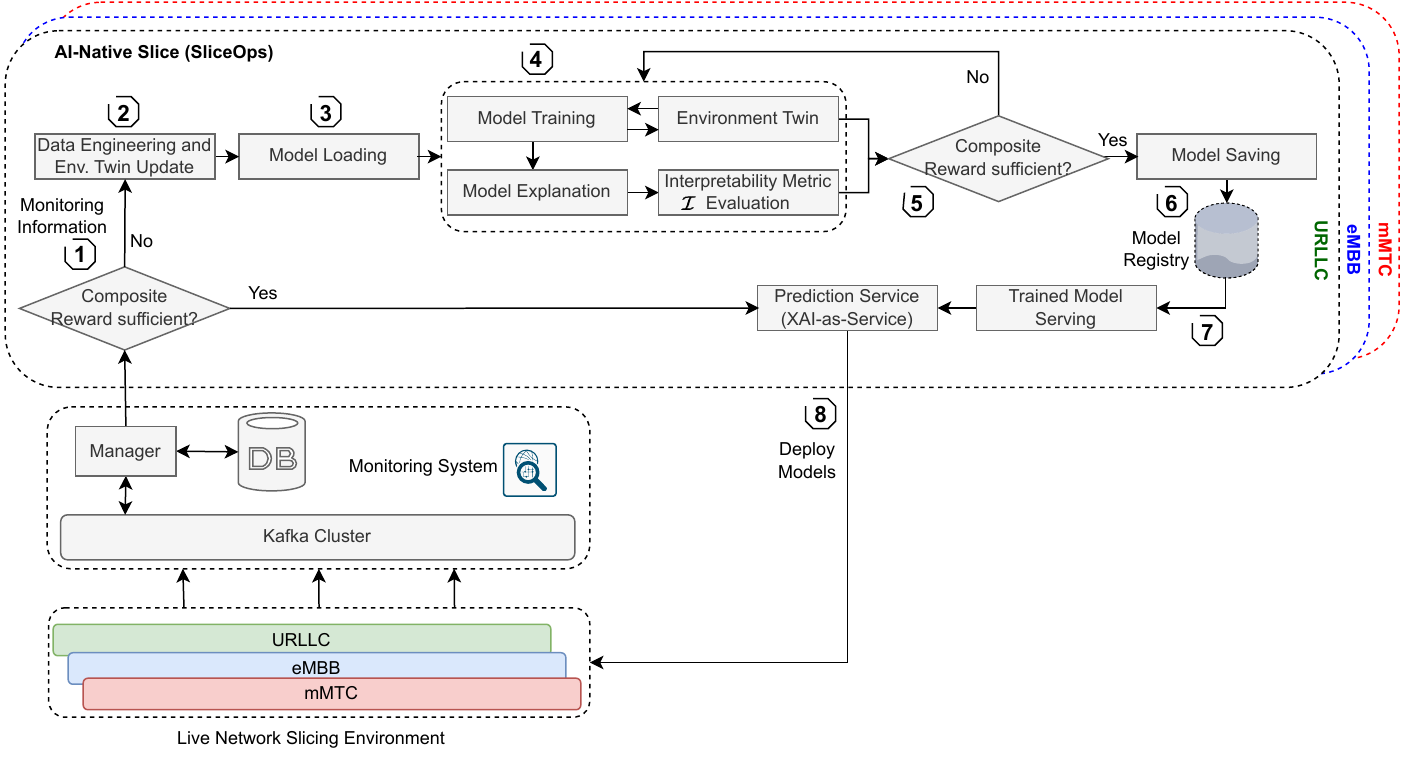}
\caption{\small The SliceOps workflow comprises of monitoring, re-training,
and deploying ML model as a service.}
\label{fig:SliceOps_architecture}
\end{figure*}

Alongside research contributions, some organizations' initiatives are relevant for MLOps components in telecommunications. The European telecommunications standards institute experiential networked intelligence (ETSI ENI) is defining a cognitive network management architecture leveraging AI techniques. The architecture aims to provide fully-automated service provision, operation, and assurance. 
The corresponding closed-loop AI mechanisms can be leveraged in the MLOps lifecycle. Unlike ETSI ENI, which focuses on AI techniques, European telecommunication standards institute zero-touch service management (ETSI ZSM) \cite{ETSI-ZSM-CL} is investigating automation challenges faced by operators and vertical industries. The group works on end-to-end (E2E) architecture and services automation to solve radical changes in the way networks are managed and orchestrated in the pivotal deployment of 5G and network slicing. 
In this intent, an MLOps approach with lifecycle management of ML models for reproducible ML pipelines would be necessary. 

There are still several issues to be solved. The stakeholders in structuring network slices need innovations and adjustments to embed MLOps into the network while enhancing the interpretability of the black-box AI decision-making process to attain human trust. To this end, we propose a revolutionary approach for 6G networks, called \emph{SliceOps} with an attention-based submodule to inspect the contribution and impact of network slice states on decision-making and choice of particular actions.

\section{Proposed Explainable MLOps Slicing}

\subsection{SliceOps Lifecycle}
\label{xai-zsm}
The SliceOps foundation relies on a practice aiming to standardize production methods through incorporating the concept of CI and CD, producing thereby reliable software and AI solutions in short cycles. In 6G AI-native networks, SliceOps modular design allows the evolution, upgrade, and scaling of the MLOps layer and its AI functions separately from the service layers.
Fig. \ref{fig:SliceOps_architecture} depicts the pictorial representation of an operationalizing AI-native slice deployed at the top of the network slicing environment with multiple SliceOps instances. The instances provide AI services to the corresponding slice (e.g., URLLC), and they can collaborate for specific tasks like resource allocation, where resources are generally shared between service slices. On the other hand, SliceOps guarantees AI performance isolation through triggering re-training, which means that if a sliceOps instance degrades for one slice, it will have minimum effect on the other slices. This is in line with containerized solutions that decouple the execution of iterative processes which is a necessity in ML pipelines. The SliceOps adheres to the below fundamental pipeline principles based on RL algorithms:
\subsubsection{Monitoring System and Data Collection (1)} It is considered as a backbone step in ML practices where data acquisition is utilized for data preparation and model design processes. It allows SliceOps agents that run as containers instances within slices to collect real-time monitoring data from the gNodeB (gNB) platform of multiple key performance indicators (KPIs), encompassing available resources, number of connected devices, bandwidth utilization, channel quality, etc. This component can store structured and unstructured
data on a very large scale. Such data is streamed through e.g., a Kafka bus to which the SliceOps AI functions can subscribe and fetch the relevant data under a specific Kafka topic name. The obtained data is used as input of the pipeline to guide the definition of policy in the form of service prediction. The SliceOps RL-based agent collects data on-the-fly (online RL) interacting with the slice environment or initializes the process with a pre-collected dataset (offline RL) stored in an internal database. The online RL can bring additional risks in terms of interacting with live environment and collected data. To solve this, we consider step 2, 6, and 8.
\subsubsection{Data Engineering and Model Loading (2, 3)} This component is responsible for data preprocessing or preparation. Different optimization targets and AI services require heterogeneous training data for neural networks. This process takes raw data from the monitoring system and transforms it into an understandable format for RL such as OpenAI Gym. The raw data contains errors and inconsistencies while having various attributes or patterns. For example, the monitored data of different slice domains can be non-Euclidean or not meeting independent and identically distributed (IID) dataset features. With the various forms of data in network slicing, pursuing the techniques such as assessment of data quality, data cleaning, data transformation, reduction of data, etc., is a vital step for better learning performance. The next step is to create neural network architecture and hyperparameters tuning before compiling and loading them into a model. The hyperparameter tuning depends highly on capability, scenario, and technology used~\cite{Globe_far}. Different datasets require setting different hyperparameters to guide the model to predict accurately in the following steps.
\subsubsection{Model Training, Evaluation and Saving (4, 5, 6)}
This module of SliceOps is segregated into a set of processes for the execution of continuous model training automatically with processed data. The ML model training runs a local optimization task, while the model explanation involves an \emph{explainer}, either attribution-based (e.g., Integrated Gradient, Saliency Maps) or perturbation-based (e.g., SHAP \cite{SHAP-ref}). Upon the evaluation of RL reward including interpretability metric, the model retraining is triggered whenever there is a model performance deterioration following new training data arrival (new inexplored states). Note that the reward should also correlate with slice targets (SLA, KPIs, etc.), while the interpretability refers to XAI metrics (attributions-based entropy, confidence, log-odds, fidelity, etc.) depending on the SLA adopted by the slice tenant. In this respect, a feedback loop between the explainer and the model is necessary to feed the model optimizer with the measured XAI metrics, enabling thereby explainability-aware learning. Finally, in case the model fulfills the target performance, it will be registered and stored in the model registry to be promoted into production later on. 
\subsubsection{AI Service Provisioning (7, 8)}
The next step after training and evaluation is to retrieve the registered model for encapsulation and move to the production stage in the form of a model prediction service. The trained model is continuously exposed and delivered to slices as representational state transfer (REST) application programming interface (API) assisted with SliceOps agents. The framework considers continuous monitoring in different steps to ensure model performance.

\begin{figure}
\centering
\includegraphics[scale=0.7]{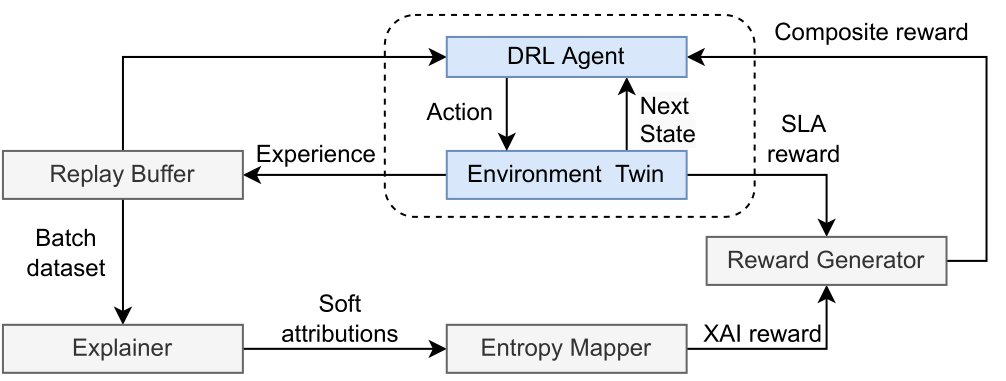}
\caption{\small The workflow of SliceOps model training based on explainable DRL.}
\label{fig:XDRL_architecture}
\end{figure}
\subsection{Explanation-Guided RL Building Blocks}

\label{XRL-description}
The aforementioned agents are assumed to be DRL-based. Unlike the conventional DRL where there is no causal relation between the input state parameters and the output action, this paper introduces an attention mechanism based on SHAP explainer, which quantifies the relevance of a state to the action and guides the XRL agent to perform explainable decisions through XAI-augmented reward shaping. The training workflow of the proposed XRL scheme is illustrated in Fig~\ref{fig:XDRL_architecture}, which is composed of the following main components, 
\subsubsection{Explainer}
It explains the DRL decision by assigning high scores (in absolute value) to the most influencing input state parameters. The score corresponds to the SHAP value, which is computed using e.g., a perturbation-based approach. Specifically, each feature is perturbed or modified while keeping other features fixed at their baseline values (e.g., white noise, zero). The model's response is observed by evaluating the perturbed instances and recording the corresponding predictions. The differences between the predictions of the perturbed instances and the baseline prediction are computed to capture the contribution of each feature when changed from the baseline value. In Sec. \ref{Evaluation-results}, we demonstrate these contributions are aggregated across different perturbations to estimate the SHAP values. Following the DRL agent interaction with the Environment Twin, it temporarily saves the experiences and observations in a replay memory/buffer which is steadily updated. Then, it generates the SHAP importance values over an extracted batch dataset of state-action.
\subsubsection{Entropy Mapper}
It applies a \emph{softmax} layer to the SHAP values provided by the \emph{Explainer} and consequently generates a probability distribution, which is  used afterward to calculate the entropy that measures uncertainty of the taken action given the input state.
\subsubsection{Composite Reward Signal}
The obtained multiplicative inverse of the maximum entropy value is used as XAI reward. In Sec. \ref{Evaluation-results}, we showcase that the composite reward---which is a sum of the SLA reward (based on meeting or violating the SLA requirements) and the XAI reward---results in minimizing the uncertainty of state-action pairs and encouraging the agent to select the best actions for specific network state values. This approach can elucidate the learning process while directing the learning toward making explainable decisions concerning a specific state.

\subsection{Benefits of SliceOps}
The main objectives of segregating the control plane from the user plane in 5G and beyond 5G networks encompass scalability, flexibility, and agility to streamline the development of new services and use cases. The proposed SliceOps approach aligns with this philosophy by creating an innovative standalone AI plane to independently provide AI services and functions to the rest of the network slices. It eliminates the necessity to modify the underlying AI functions and structure of network slice instances for new services. By a decoupled AI plane from the control plane and user plane, the network can upgrade or add new AI and automation functions without impacting other planes. This flexibility and mentioned explainability features of SliceOps pave the way for faster deployment and trustworthy network optimizations in network slicing.

\subsection{ETSI ZSM Compliance}
The design of SliceOps solution closely adheres to the ZSM framework. The standardization process for ETSI ZSM is still in its early stages, with preliminary specifications based on a high level of abstraction. The core concept of the closed-loop AI system is to utilize context-aware and metadata-driven policies to more efficiently and quickly identify and incorporate new conditions while updating knowledge and making robust and actionable decisions. As shown in Fig. \ref{fig:SliceOps_architecture}, SliceOps is a practice toward deploying a more realistic closed-loop scheme to fulfill viable automation solutions for network slicing control. SliceOps manages the lifecycle of AI models in a closed-loop way that includes model performance monitoring, re-training, and delivery. It extends the ZSM framework to manage, besides the service functions, the underlying AI functions, which are natively supported in a standalone slice. Moreover, the SliceOps agents are based on the closed-loop workflow in Fig. \ref{fig:XDRL_architecture}.

\subsection{O-RAN Slicing Compliance}
\label{sec:o-ran}

The SliceOps framework can be adapted to use case 3 of open radio access network (O-RAN) slicing, which is related to resource allocation optimization, requirements, and architecture \cite{o-ran-slicing}. Specifically, SliceOps layer can be developed as rApps at the non-real-time (Non-RT) radio intelligent controller (RIC). Thanks to its architecture that brings valuable differentiators by leveraging XAI and CI/CD, SliceOps would implement new use cases with agility and automate network operations. In this respect, rApps need to access abundant monitoring data---such as traffic load, latency, and signal strength---through the O1 interface to efficiently carry out their designated functions in time-demanding training procedures. Then, the model or policy trained by SliceOps can be packaged as an artifact and delivered via the A1 interface to run on the Near-RT RIC interface as xApp. This model deployment is performed using REST APIs over HTTP for the transfer of such JSON objects. For dynamic network optimization purposes, this xApp would control the underlying O-RAN components, namely, the central unit-control plane (O-CU-CP), different slice open central unit-user plane (O-CU-UP), and open distributed unit plane (O-DU) by using the E2 interface. The E2 interface encompasses two protocols, namely the E2 application protocol (E2AP) and the E2 service model (E2SM).

\section{Evaluation}
The next important step after introducing the key elements of the proposed framework is to exemplify the SliceOps approach and evaluate the performance to verify the benefits of XAI in the MLOps pipeline. In this section, we demonstrate the implementation of a small-scale SliceOps framework for network slicing and then the effectiveness of the model is validated in terms of ameliorating the long-term revenue (average reward), transmission latency, and dropped traffic.


\subsection{Network Architecture and Experiment Parameters}
\label{server-spec}
For the sake of validating the SliceOps framework in realistic settings, we consider a gNB scenario, wherein a set of slices $\mathcal{I}$ is deployed. The scenario includes $three$ slices, i.e., URLLC, enhanced mobile broadband (eMBB), and massive machine-type communications (mMTC). The slices are characterized by the SLA latency $\Lambda_i = [10, 40, 20]$ ms, respectively. 
Without loss of generality, the considered gNB is characterized by the radio capacity $C=100$ physical resource blocks (PRBs) of a fixed bandwidth and assumes the slices running over gNB simultaneously. The slice traffic demand is modeled as the realization of a Poisson distribution with mean value $\lambda_i$ and emulates the signal-to-noise ratio (SNR) variability extracting its instantaneous values from a Rayleigh distribution with the average value set to $25$ dB. We set $\iota = 10$ PRBs as the minimum resource allocation step. The framework leverages Python programming language, exploiting OpenAI Gym library~\cite{Globe_far} and interfacing DRL agents with a custom gNB simulator environment \cite{Specialization_TVT}. The simulator consists of virtual transmission queues and main physical (PHY), medium access control (MAC), and radio link control (RLC) functionalities. 
Each SlicOps agent is endowed with a double deep Q-network (DDQN) \cite{GAN_Powered}. The agents interact with each other and 
network environment to gather the slice networking statistics (e.g., channel quality, served traffic, consumed resources, etc.). Then, they enforce PRB policy decisions provided by the corresponding SliceOps layer in the gNB slice scheduler. We use a dedicated server equipped with two Intel(R) Xeon(R) Gold 5218 CPUs @ 2.30GHz, two NVIDIA GeForce RTX 2080 Ti GPUs, and the DNNs are implemented based on TensorFlow-GPU version 2.5.0. The neural network architecture uses two fully connected layers with \emph{24} neurons activated by ReLU function. The network parameters are updated using the Adam optimizer. The discount factor $\gamma$ and learning rate $\xi$ are set to be \emph{0.99} and \emph{0.001}, respectively. The replay buffer size of each agent $\beta_{i}$ is \emph{20000} samples, out of which a batch of \emph{32} samples is extracted for each training interval. To deploy the solution in the cloud-native mode, we leveraged a containerized approach where a cloud server hosts SliceOps instances and corresponding modules responsible for providing AI service for different problems in slices. On the other hand, SliceOps agents of slices run by using the Docker compose tool and communicate with the server through FastAPI as a REST API.

The operational training phases are discussed in Sec.~\ref{XRL-description}.




\subsection{Latency-Aware Resource Allocation}
\label{Evaluation-results}
Acquiring swift and constructive resource allocation in network slicing is precluded due to the lack of dynamic traffic steering. We cast the radio resource allocation problem in gNB as an optimization problem, emphasizing on minimizing allocated resources and latency to meet the SLA. We consider the transmission latency as the average time that traffic of a slice experiences before being served within the gNB transmission buffers due to the inter-slice scheduling process. The radio resource availability for the downlink traffic is divided into subsets of PRBs. The provided AI model by SliceOps instances lets the SliceOps agent dynamically assign the optimum PRBs to each network slice following the real-time traffic and SLA requirements. In this scenario, we consider correct and fair dimensioning of the inter-slice PRB enforcement instead of focusing on the intra-slice scheduling issue.  
\begin{figure}[h]
\centering
\includegraphics[scale=0.55]{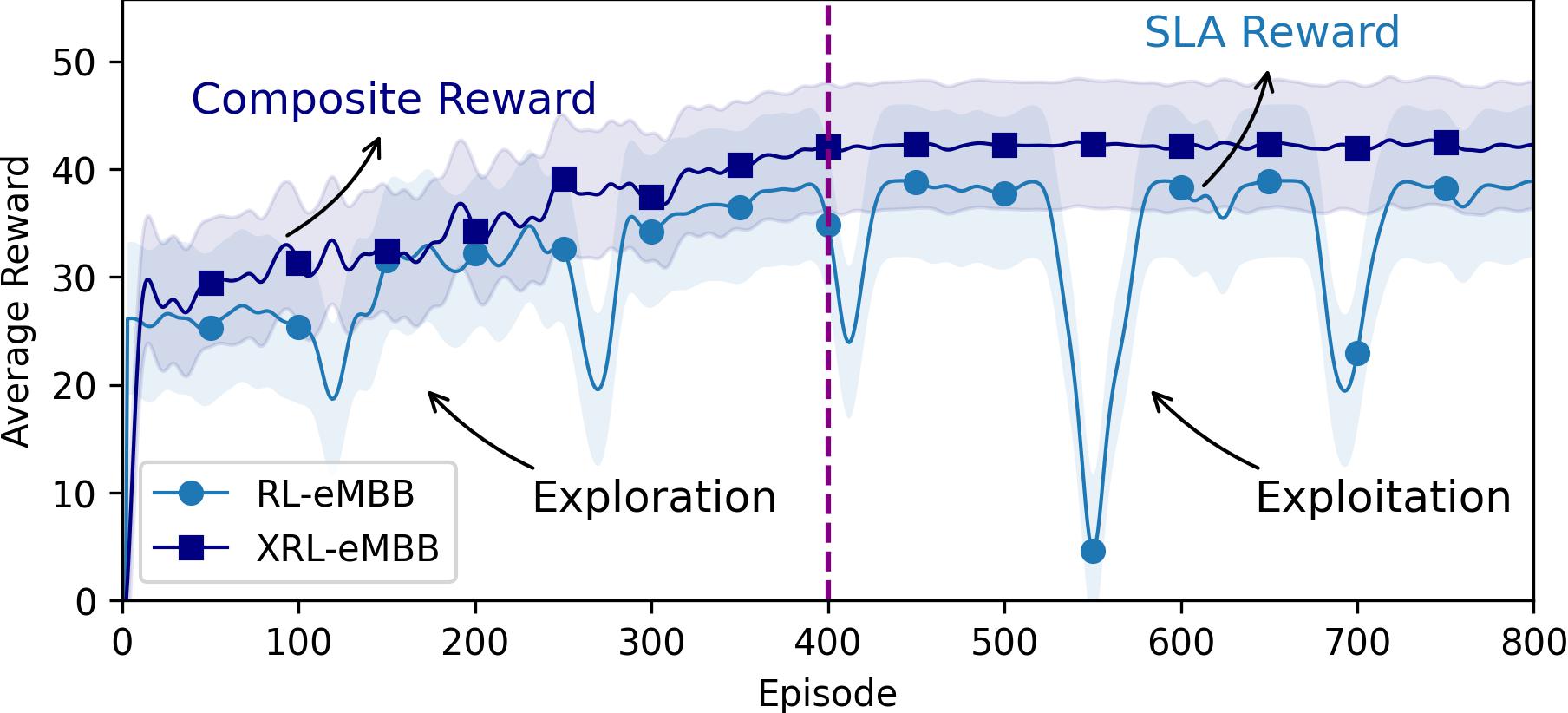}
\caption{\small Convergence performance of the RL and XRL approaches. For the sake of visual clarity, the curves are smoothed concerning confidence bands and standard deviation.}
\label{fig:rl-xrl-reward}
\end{figure}

As shown in Fig.~\ref{fig:rl-xrl-reward}, the designed composite reward function (XRL strategy) assisted by the SHAP approach guarantees better learning generalization and robust performance compared to the RL method (SLA reward) as the baseline. In around half of the overall training, the eMBB SliceOps agent inspects action space (PRBs) that initially leads to high fluctuations in learning curves, i.e., exploration, and then strives to achieve the right trade-off between the learned decision policies and varying network states, i.e., exploitation. It is safe to assert that the composite reward based on the proposed explanation-guided action-selection strategy enhances the performance of SliceOps agents. The SHAP explainer extracts the features and their importance values from the batch dataset for a particular prediction in conjunction with the entropy mapper to provide a reward metric to guide the agent with more relevant state-action pairs.

\begin{figure}[h]
\centering
\includegraphics[scale=0.5]{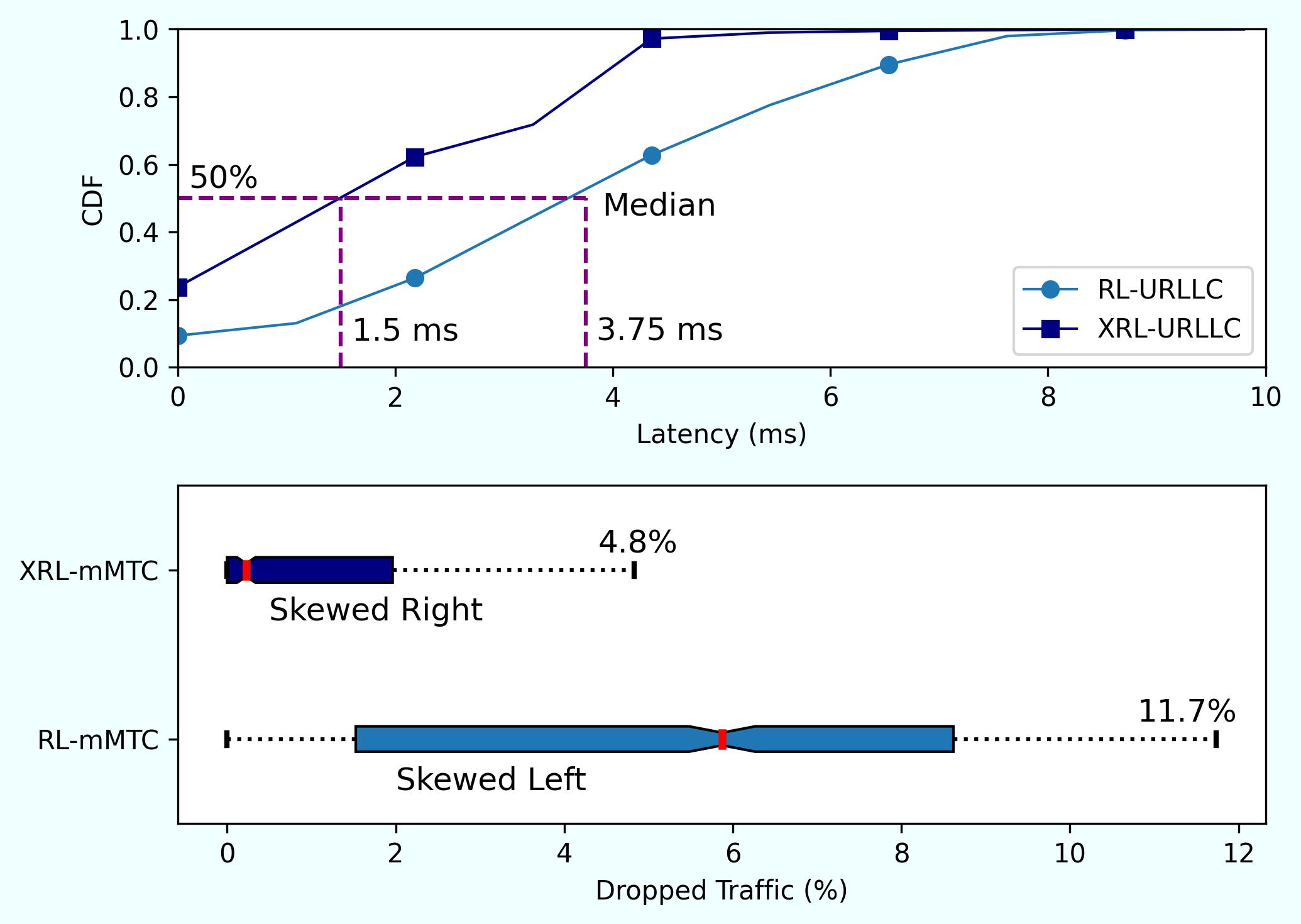}
\caption{\small Network performance comparison for RL and XRL settings. (Up) The  transmission latency CDF for URLLC slice, (Down) The performance evaluation in terms of dropped traffic for mMTC slice.}
\label{fig:rl-xrl-Latency}
\end{figure}

Fig.~\ref{fig:rl-xrl-Latency}~(Up) depicts the cumulative distribution function (CDF) of the time that URLLC traffic experiences within the gNB transmission buffer, resulting from RL and XRL-based SliceOps resource allocation. From the results, it can be noticed that the XRL approach leads to higher performance where \emph{50\%} of perceived latencies in the URLLC slice is less than \emph{1.5~ms}, whereas this value for the RL solution is \emph{3.75~ms}. The performance of XRL-SliceOps agents reveals that they allocate adequate radio resources proportional to the traffic demand while handling varying resource contention among slices. In contrast, the defective collaboration among agents in the RL strategy and the erroneous trade-off between resource allocation and network conditions (state-action pairs) result in higher incurred latency.

We continue the performance analysis on the proposed XRL approach by shedding light on the volume of dropped traffic that does not meet latency SLA requirements owing to mistaken radio resource allocation policies, as showcased by Fig.~\ref{fig:rl-xrl-Latency}~(Down). The box plot illustrates the highest value of mMTC dropped traffic, excluding outliers for the XRL scheme, is \emph{4.8\%} whereas this value for the RL method is \emph{11.7\%}. Besides, the lopsided box plot of XRL is positively skewed where the mean value is greater than the median, i.e., the majority of the values is located on the left side (lower dropped traffic values). In contrast, mMTC slice experiences higher dropped traffic by RL solution where the box plot indicates a few exceptionally small dropped traffic and most values are large, which results in the mean being pulled to the left.

\begin{figure}[h]
\centering
\includegraphics[scale=0.55]{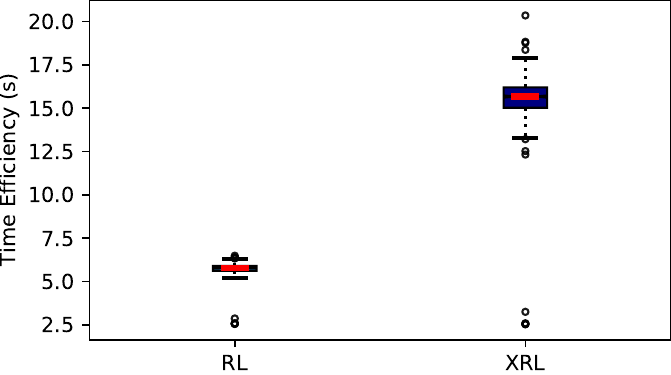}
\caption{\small Comparison of algorithms in terms of time efficiency on the network slicing setup. All evaluations were conducted on the defined server in Sec.~\ref{server-spec}}
\label{fig:rl-xrl-time}
\end{figure}

Fig.~\ref{fig:rl-xrl-time} reflects time efficiency comparison between RL and XRL algorithms during the whole training operation. The box plot illustrates that the median line of the box for RL and XRL is at \emph{6s} and \emph{16s}, respectively. At a glance, we can explicate that the RL algorithm takes less processing time to complete episodes than XRL. However, as shown in Fig.~\ref{fig:rl-xrl-reward}, XRL converges faster, which can compensate this higher complexity compared to the RL approach. Note that there are some outliers for each box between \emph{2.5s} and \emph{3s}, which is the period of filling the replay buffer initially as a part of training different DRL algorithms.
\begin{figure}[h]
\centering
\includegraphics[scale=0.46]{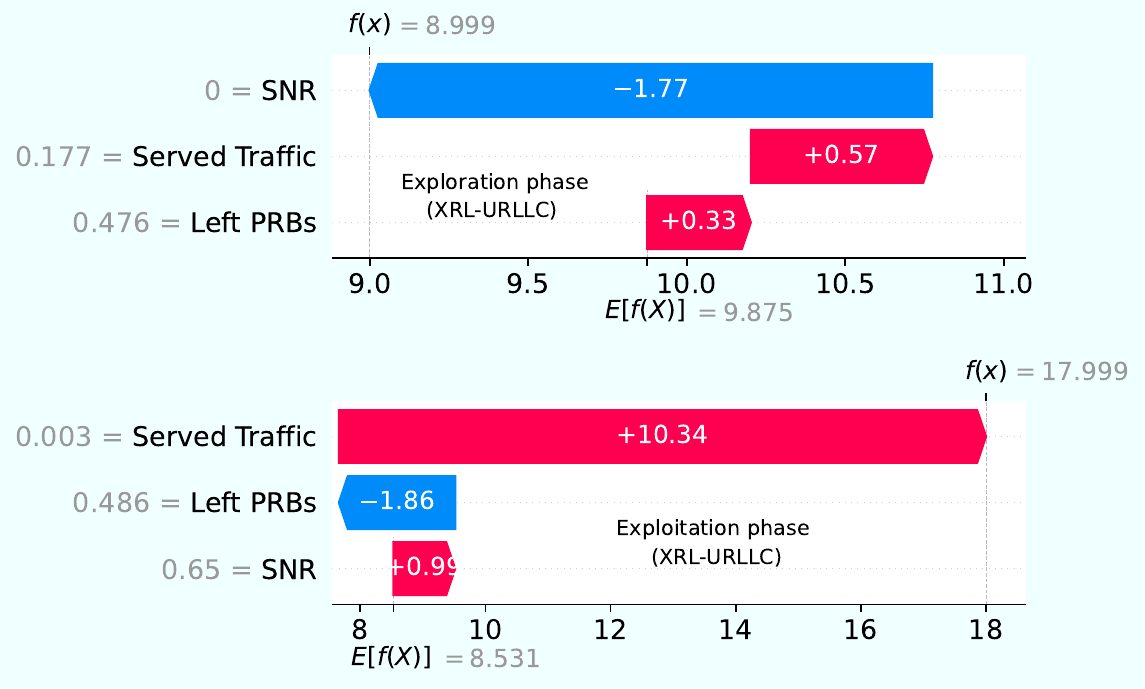}
\caption{\small The XAI waterfall plot illustrates the contribution (either positive or negative) of a given input state parameter to the output decision for URLLC slice. (Up) Exploration phase, (Down) Exploitation phase.}
\label{fig:SHAP-values}
\end{figure}

In Fig.~\ref{fig:SHAP-values}, we observe that $f(x)$ represents the predicted action taken by the XRL agent, which involves allocating PRBs to the URLLC slice. Meanwhile, $\mathbf{E}[f(x)]$ represents the expected value or the average of all possible actions. The absolute SHAP value provides insights into the influence of a single state on the action taken. During the initial stages of training, the agent behaves as a Max C/I scheduler, resulting in a penalty for URLLC users experiencing a low SNR state (SHAP value $=-1.77$). Consequently, this leads to a relatively low allocation of PRBs per slice, specifically $8.99$ PRBs. Contrastingly, in episode $500$, which signifies the exploitation phase where the agent has learned the optimal policy, the agent's action primarily depends on the served traffic (SHAP value $=10.34$). As a result, a higher allocation of PRBs, specifically $17.99$ PRBs, is observed. In Fig.~\ref{fig:SHAP-values} (Up), the exploration entails engaging in actions with uncertain PRBs allocation policy aims to acquire deep insights about the network environment to learn better optimal strategies. Fig.~\ref{fig:SHAP-values} (Down) presents the results in exploitation phase where agent leverages the learned knowledge in exploration phase and executing PRB actions that are estimated to yield the highest rewards. When employing various background batch sizes obtained through random sampling, the SHAP values and variable rankings display fluctuations, which impedes their trustworthiness. Fortunately, as the size of the background batch
increases, these fluctuations tend to diminish.


\section{Challenges and Limitations of SliceOps}
\subsection{Time-Sensitive Slicing}
MLOps can speed up the different steps of AI operations in slices, such as data collection and preparation process, and provide faster response to changed conditions of the network. However, it suffers from the extensive process and long time E2E procedure.
Thus, optimizing the training time and ameliorating inference model performance is necessary to harness the full potential of MLOps into a slice instance.

\subsection{Secure Slicing}

Network slicing may introduce new vulnerabilities, while the stringent performance of slices in a shared infrastructure environment requires a secure E2E network. 
In this intent, slice isolation play a vital role in guaranteeing safe and accurate operations. 
The secure data pipelines in MLOps are challenging and need to consistently set governance rules in each pipeline step to protect against different data attacks.

\subsection{Collaborative Slicing}

The learning procedure for establishing large-scale intelligent network slicing is not trivial since it is time-consuming and computationally complex, especially if each slice is independently trained for the same tasks. By relying on transfer learning with a collaborative approach between slice agents, a partially trained AI model can be distributed to different slices with similar tasks. Nonetheless, the sheer number of model transfer can raise privacy and isolation concerns.

\section{Future Research Directions}
\subsection{Sustainability}

In 6G, we will witness a massive number of slices that deal with significantly more data at faster rates than the current network's deployment. It is of utmost importance to derive more suitable power-optimized solutions to ensure the long-term sustainability of AI-driven slicing.
Therefore, more research should be conducted to reduce computations and energy consumption in different steps of MLOps-driven slicing.

\subsection{SliceOps-Enabling O-RAN}
We discussed the compliance of SliceOps in O-RAN slicing architecture \cite{o-ran-slicing} in Sec.~\ref{sec:o-ran}. Nevertheless, It represents a considerable challenge due to the lack of principles and distributed architecture to underpin AI-driven massive 6G slicing. It is an exciting line of research to achieve reproducible AI model deployment in an O-RAN massive slicing setup.

\subsection{SliceOps-Enabling Metaverse Services}

In the ambitious vision of 6G, the coined xURLLC will become conflated with both eMBB and mMTC.  
The xURLLC paves the way for the implementation of immersive and Metaverse services. 
In this intent, the feasible MLOps solutions need more study and analysis to satisfy requirements of xURLLC services, such as latency and bandwidth.

\section{Conclusion}
Algorithmic and architectural innovations are required to streamline 6G slicing automation in future networks. This paper has introduced SliceOps, a framework for AI-native 6G networks, where the AI operations (MLOps) are gathered in a standalone slice that provides AI service to the rest of slices. This AI slice extends the ZSM closed-loop to the AI lifecycle management. Moreover, explainability-guided learning is adopted in SliceOps to ensure trust in and robustness of the DRL agents. Both AI and network performance results underpin the proposed framework. Finally, the challenges and future research directions are identified.


\end{document}